# 基于双摆系统的 Astrojax 双球轨迹的研究与实验设计

段 彬,白子豪,张宇隆,张清源,房思行,邵勃蕙*

(吉林大学 物理学院,长春 130022)

**摘 要**:基于双摆系统,结合拉格朗日方程在保守系统下进行分析,通过双目测定三维捕捉装置对运动质点进行捕捉,建立了 Astrojax 的两种轨迹模型,并给出轨迹经验公式、参数之间的关系。通过研究,得到轨迹公式及相关参数之间的理论数据与实验数据拟合良好,可准确的预测模型的运动轨迹并可使小球运动轨迹有针对地改变。该实验所需设备及材料均简单易得,实验主题有趣且新颖,可作为大学物理实验课程中的拓展实验,使学生了解双摆系统的运动特性,从生活中发现物理、学习物理。不仅能够提高学生学习兴趣,还可拓宽学生的知识面与培养学生的动手能力。

**关键词**: 双摆系统;拉格朗日方程;Astrojax 双球;实验教学

**中图分类号**:G 642;O311.2    **文献标识码**:

## Research and experimental design of Astrojax double balls trajectory based on double pendulum system

DUAN Bin, BAI Zihao, ZHANG Yulong, ZHANG Qingyuan, FANG Sixing, SHAO Bohui*

(College of Physics, Jilin University, Changchun 130022, China)

**Abstract**: Based on the double pendulum and Lagrange equation, the moving particles are captured by a binocular three-dimensional capture camera. Two trajectory models of Astrojax and the relationship between trajectory empirical formula and parameters are established. Through research, the calculated trajectory of this formula and related parameters fit well with the actual measured trajectory, and can accurately predict and change the trajectory of the model. The equipment and materials required in the experiment are simple and easy to obtain, and the experimental theme is relatively interesting and novel, which can be applied as an extended experiment in college physics experiment course, so that students can understand the motion characteristics of the double pendulum and learn physics from life. The designing experiment can not only improve students' interest in learning, but also broaden their knowledge and cultivate their practical ability.

Astrojax[1]是由 Larry Shaw 于 1986 年发明的一种玩具，在世界上大受欢迎。它由三个球组成，两个球固定在绳的两端，中心球可在两个端球之间沿线自由滑动。2000 年，Jerrold E. Marsden[2]等人利用集合力学和对称分岔理论研究双摆的一些动力学特征，为进一步研究 Astrojax 小球提供了理论基础。2005 年，Philip Du Toit[3]提出 Astrojax 的物理结构本质上是一个双球面摆，其中第一个摆锤可以在绳上自由滑动，根据拉格朗日约化理论、变分积分和唤醒理论来研究 Astrojax 摆。2014 年，Andras Karsai[4]等人研究了广义双摆的性质，关注摆绳长度动态变化，分析了轨迹的平均混沌寿命和主振荡频率，发现该混沌摆的振荡频率与低频区的强迫频率线性相关，并得出结论说明没有反馈机制系统的、一个天然的周期性强迫并不足以创造 Astrojax 摆稳定的、长寿命的轨道。本文从实验设计及实验数据测量出发，手动给予系统反馈机制。通过周期性地控制绳子自由端的方式调节小球运动，分析小球的运动轨迹。实验表明，通过绳子自由端的不同周期性运动，会使小球产生两类不同的运动形式，并且每种运动情况下运动轨迹与两小球的质量紧密联系。

本文通过自行搭建实验装置,结合双目测定三维捕捉装置捕捉绳子的自由端以及两个小球的轨迹，建立起两种小球的相对运动的物理模型。通过 MATLAB 以及 Mathematics 将理论计算轨迹与实际观测轨迹相拟合，验证公式正确性。实验设计思路简明、仪器耗材易得、实验结果重复性高，可应用于大学物理实验教学中，使学生了解双摆系统的特性，提高学生的物理实验能力。

## 1 实验设计及测试方法

平面双摆系统由两个单摆连接而成，通过无质量刚性杆连接小球。对其传统的动力学分析仅限于在摆幅很小的条件下的振动系统、阻尼系统和受迫振动。而本实验是一个复杂的非常规的双摆系统，如图 1 所示。在非线性的动力学特征之下，系统额外增加大振幅、多自由度和参数激励更使其易处于无规则的状态。

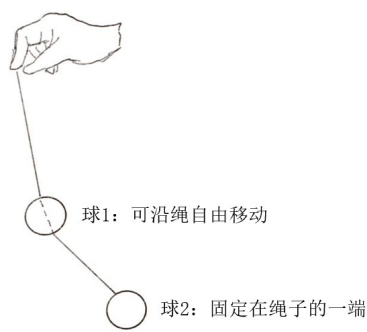

图 1 实验设计图

通过大量的实验发现，当绳子自由端有竖直方向运动时，机器的严格周期激励反而使小球轨迹在更快的时间内发散甚至混沌。而当人手闭环反馈激励时，小球的轨迹呈现类周期的状态。这是因为双摆系统具有发散性和混沌性，它对初始条件极为敏感。在机器激励的情况下，环境微扰导致轨道发散。相比之下，人可以通过眼睛观测实现反馈调节，对发散轨道产生负反馈作用，从而使得在一定时间内系统呈现类周期的运动。因此，为保证双摆系统能保持一定条件的稳定性，本文的在有竖直运动的实验时均采用人手动改变绳子自由端。

由于本实验的小球运动为三维空间的运动，且在人手调节时无法使用固定公式进行表达，系统很难用普通的相机进行拍摄三维运动。因此本文采用双目测定三维捕捉装置，如图二所示，该装置模拟人眼对深度的感知，利用视差测距的方式，采用多台高速摄像机，得到相机

的内参以及相对位置等外参进行标定，实现高速球体三维位置测定。为避免可见光的影响，我们选择了特殊频率的红外线进行照射。在物体上贴有反光圆片，通过多个相机的捕捉，利用视差，判断小球圆心的位置，从而得到三维坐标。这套装置的精度可以达到空间误差在 1 毫米以内，时间误差在 5 毫秒以内。

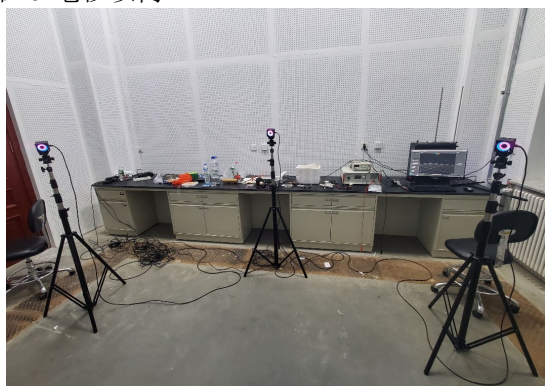

图 2 三维捕捉装置示意图

通过双目测定三维捕捉装置对运动轨迹进行捕捉，得到了如下两种模式：

滞后摆，如图 3 所示：当机器在水平方向上旋转自由端做平面圆周运动，可得两小球在在水平相位上进行稳定运行，且相位差相差 π；

天体摆，如图 4 所示：当手动上下移动绳子自由端时，会发现两小球出现类似于天体之间，一个围绕另一个球旋转的运动，因此被命名为天体摆。

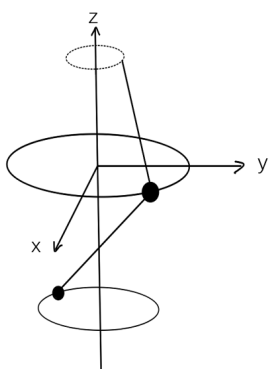 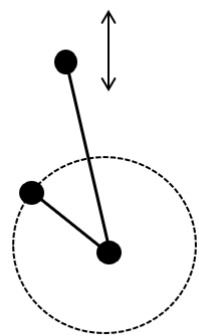

图 3　滞后摆示意图　　　　　　　图 4　天体摆示意图

## 2　理论分析

### 2.1　滞后摆；

如图 5 所示，对于整个系统存在五个广义自由度 $\lambda, \theta_1, \theta_2, \varphi_1, \varphi_2$。

假设普适下周期性移动自由端悬点为 ($\varepsilon x(t), \varepsilon y(t), \varepsilon z(t)$)

用广义坐标表示两小球质点的笛卡尔坐标：

$A(x_1, y_1, z_1)$　　　　$B(x_2, y_2, z_2)$

系统动能表达式为：

$$T = \frac{1}{2}m_1\sqrt{\dot{x}_1^2 + \dot{y}_1^2 + \dot{z}_1^2} + \frac{1}{2}m_2\sqrt{\dot{x}_2^2 + \dot{y}_2^2 + \dot{z}_2^2} \tag{1}$$

系统势能表达式为：$U = m_1 g z_1 + m_2 g z_2$ （2）

由拉格朗日函数：$L = T - U$ （3）

建立保守系统下的五个自由度的拉格朗日方程[3]：

$$\frac{d}{dt}\left(\frac{\partial L}{\partial \dot\lambda}\right) - \frac{\partial L}{\partial \lambda} = 0, \quad \frac{d}{dt}\left(\frac{\partial L}{\partial \dot\theta_1}\right) - \frac{\partial L}{\partial \theta_1} = 0, \quad \frac{d}{dt}\left(\frac{\partial L}{\partial \dot\theta_2}\right) - \frac{\partial L}{\partial \theta_2} = 0,$$

$$\frac{d}{dt}\left(\frac{\partial L}{\partial \dot\varphi_1}\right) - \frac{\partial L}{\partial \varphi_1} = 0, \quad \frac{d}{dt}\left(\frac{\partial L}{\partial \dot\varphi_2}\right) - \frac{\partial L}{\partial \varphi_2} = 0 \quad (4)$$

这里我们研究系统稳定运行过程，即两小球在各自的平面圆形轨道下做稳定运动。

假设自由端为圆周轨迹为

$$\begin{cases} \varepsilon_x(t) = \gamma \cos\beta t \\ \varepsilon_y(t) = r \sin\beta t \end{cases} \quad (5)$$

式中 r 为偏心距，$\beta$ 为驱动角频率，z 方向无驱动。

当系统保持相对稳定时，各角度满足条件；

$$\begin{cases} \dot\phi_1 = \dot\phi_2 = \omega \\ \dot\lambda = \ddot\lambda = 0 \\ \ddot\theta_1 = \ddot\theta_2 = \dot\theta_1 = \dot\theta_2 = 0 \\ \ddot\phi_1 = \ddot\phi_2 = 0 \end{cases} \quad (6)$$

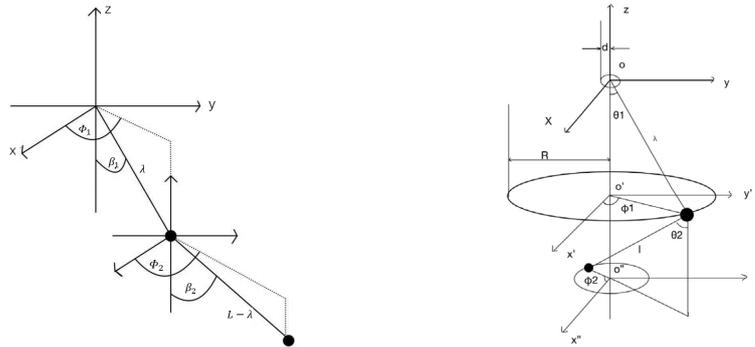

图 5  滞后摆角度示意图

通过稳定时各角度满足的条件（6）化简五个拉格朗日方程，可得到以下结果：

$$\begin{cases} \tan\phi_1 = \dfrac{-\varepsilon_x'}{\varepsilon_y'} \\ \tan\phi_2 = \dfrac{-\varepsilon_x'}{\varepsilon_y'} \\ \lambda = \dfrac{m_2 l \omega^2 \cos\theta_1 \sin\theta_2 + (m_1+m_2)g \sin\theta_1}{\frac{1}{2}(m_1+m_2)\omega^2 \sin 2\theta_1 + m_2 \omega^2 \cos\theta_1 \sin\theta_2} \end{cases} \quad (7)$$

其中：(1)(2) 验证了 $\Phi_1 = \Phi_2 + \pi$，及两小球相位上的滞后现象。

(3) 式为评估式，及通过实验验证左右两端是否一致。

## 2.2  天体摆；

对于竖直轨道模式，从实验上发现：
1. 中间小球的轨迹近似为直线运动
2. 固定小球轨道为椭圆轨道，长短半轴之比非常接近 2：1
3. 手的振动与小球振幅相比为小量
4. 球 1 与球 2 之间的距离（即绳长）几乎不变。

基于上述四个实验事实，本文将竖直轨道模式简化为在$XOY$平面的平面摆模型。

如图 6 所示，模型可以简化为：中间小球在杆上运动，末端固定另一个小球，整个系统在水平方向初始角动量为 0。运动开始时给系统一个初能量，使系统成为质量为$m_2$的平面摆，其悬挂点$m_1$在$m_2$所在的运动平面做往复的直线运动

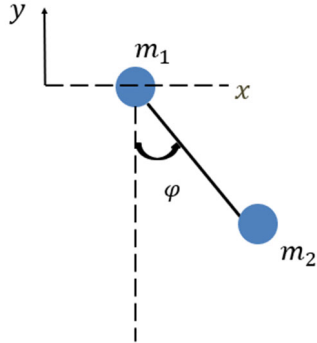

图 6　简化后平面方向天体摆示意图

建立拉格朗日方程：

$$L = \frac{1}{2}(m_1 + m_2)\dot{x}^2 + \frac{1}{2}m_2(l^2\dot{\varphi}^2 + 2l\dot{x}\dot{\varphi}\cos\varphi) + m_2 gl\cos\varphi \qquad （8）$$

$$\frac{\mathrm{d}}{\mathrm{d}t}\left(\frac{\partial L}{\partial \dot{x}}\right) - \frac{\partial L}{\partial x} = 0 \qquad （9）$$

$$\frac{\mathrm{d}}{\mathrm{d}t}\left(\frac{\partial L}{\partial \dot{\varphi}}\right) - \frac{\partial L}{\partial \varphi} = 0 \qquad （10）$$

同时建立广义坐标微分方程组：

$$m_2 l^2\ddot{\varphi} + md\ddot{x}\cos\varphi + m_2 g_l \sin\varphi = 0 \qquad （11）$$

$$(m1 + m_2)\ddot{x} + m_2 l\ddot{\varphi}\cos\varphi - m_2 l\dot{\varphi}^2 \sin\varphi = 0 \qquad （12）$$

使用 Mathematica 进行数值模拟，可得到小球轨迹以及变量对运动轨迹的影响。

## 3　小球运动轨迹测定实验及结果

### 3.1　滞后摆；

随机选用材料参数：$m_1 = 0.004\ 8\ \text{kg}$、$m_2 = 0.005\ 4\ \text{kg}$、$L = 0.6\ \text{m}$

因系统为平面二维运动，机器可带来一定的稳定性，因此为减少部分工作量本文换用不同转速(1,2,3)即（240、270、350 rpm）档风扇，在两小球稳定运行时，通过视差测距三维坐标捕捉进行角度与长度的测量（测十组，取其平均值），带入公式（7）。

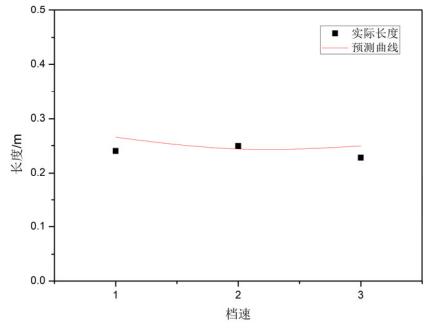

**图 7 λ实际长度和理论长度的拟合曲线**

如图7所示,实际长度与理论长度数据拟合良好,得到的数据与理论误差均在10%以内,因此在滞后摆中理论与实验进行了一个很好的验证。

### 3.2 天体摆；

1）利用公式（11）、（12）通过 Mathematica 进行数值模拟，针对球 2 的运动进行模拟并和实际轨道对比，可得到理论球 2 的移动轨迹如图 8 所示。

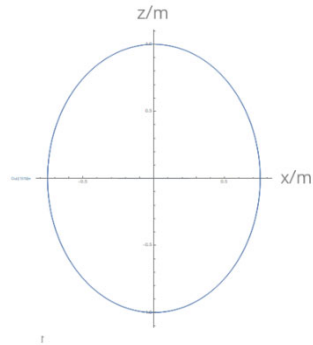

**图 8 理论得出球 2 的移动轨迹**

通过数值多次模拟以及实验相对比，发现球 2 的运动轨迹在 z 轴方向始终是 x 轴方向的二倍。

结论 1：球 2 的轨道是长短半轴 2：1 的椭圆

2）当选用的两球质量相同时,通过模拟可得到运动轨迹在 X 轴、Z 轴方向随时间的变化情况如图 9 所示。

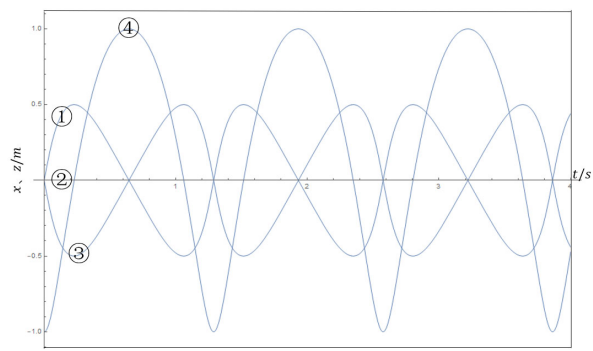

**图 9  运动轨迹随时间的变化情况**

其中是①球$m_1$的 x 轴的轨迹，②球$m_2$的 z 轴轨迹，坐标恒为 0，③为固定小球的 x 轴的轨迹，④为固定小球的 z 轴。

通过多次仿真可得知无论是形状还是周期都与质量无关，即 0.1kg 和 100kg 的小球，他们的所有参量都是一样的。

固定$\frac{m_1}{m_2}$=1 取随机尺寸、质量的小球以及随机的绳长的轨道，进行理论实验对比。进行 22 组数据，通过增加一个广义力将得到的周期进行能量修正。

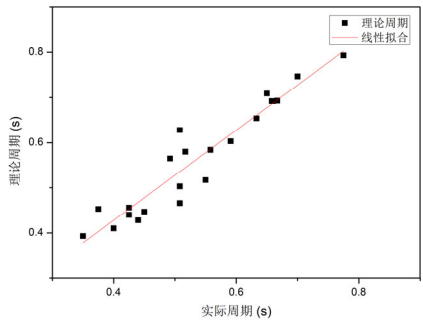

**图 10  实际周期和理论周期的关系**

如图 10 所示，通过能量修正后的实际周期和理论周期偏差很小。11 组实验偏差率小于 5%，17 组实验偏差率小于 10%。线性拟合后$R^2 = 0.906\,6$，平均偏差为 5.5%，实验与理论数据拟合较好。

结论 2：数值轨道形状与周期与小球质量 m 无关

3）改变两小球质量比例，当质量比例由无穷小到无穷大，轨道横纵比 f 、球 2 振幅与绳长比值 a 与$\frac{m_2}{m_1}$的关系通过 Matlab 进行数据模拟。

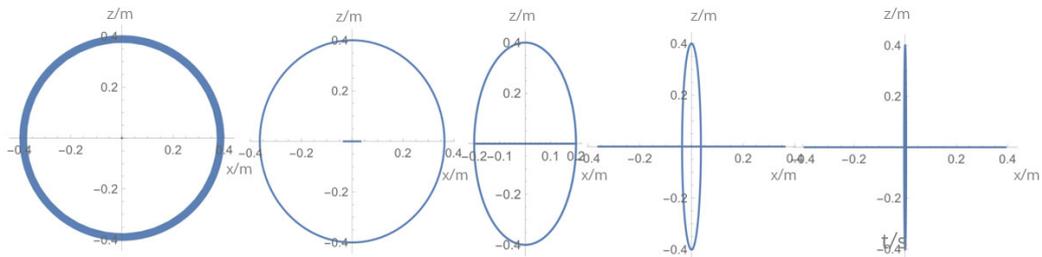

**图 11  球 2 单一方向的轨迹变化**

如图 11 所示，球 2 轨道从半径为 1 的圆逐渐变为纵向半轴为 1、横向越来越扁的椭圆；球 1 轨迹从一个点逐渐变为长度为 1 的一条线段；当$\frac{m_2}{m_1}$= 1 时，固体球横纵比为$\frac{1}{2}$，两小球横

向振幅相同。可以得出，轨道横纵比 f、球 2 振幅与绳长比值 a 与质量比例 $\frac{m_2}{m_1}$ 有关。

通过 Mathematic 进一步的数值模拟探究轨道横纵比 f、振幅与绳长比值 a 与质量比 $\frac{m_2}{m_1}$ 的变化规律如图 12、13 所示。

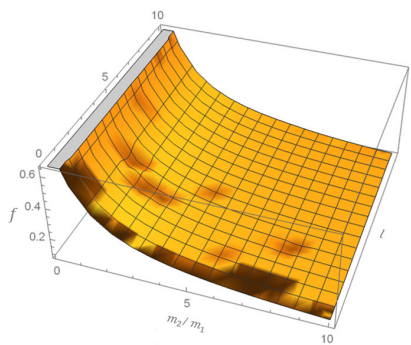 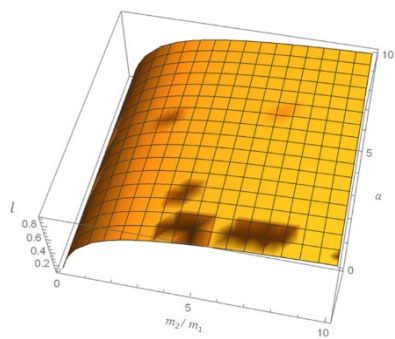

图 12　轨道横纵比 f 的变化　　　　图 13　振幅与绳长比值 a 的变化

如图 14、15 所示，固定 $\frac{m_1}{m_2}$= 0.35 取随机尺寸、质量的小球以及随机的绳长的轨道，进行理论实验对比；固定 $\frac{m_1}{m_2}$= 2.87 取随机尺寸、质量的小球以及随机的绳长的轨道，进行理论实验对比。

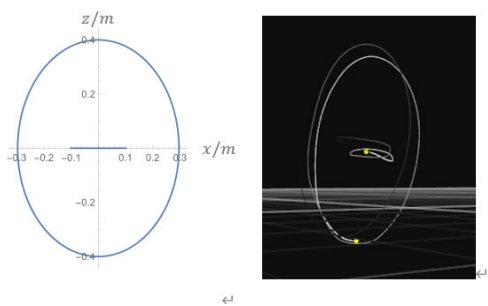 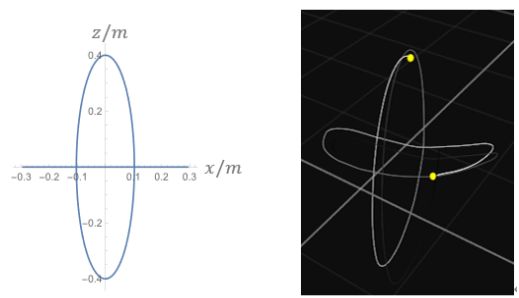

图 14　当 $\frac{m_1}{m_2}$=0.35 时 仿真与实验的对比　　　图 15　当 $\frac{m_1}{m_2}$=2.87 时 仿真与实验的对比

通过综上三种模拟和实验对比，可以观察到模拟结果和实验结果吻合良好。

结论 3：轨道横纵比 f、球 2 振幅与绳长比值 a 与质量比例 $\frac{m_2}{m_1}$ 有关，与绳长 $l$ 无关。

4）针对轨道周期进行研究，通过微分方程组（11）、（12）进行 Mathematic 微分方程数值模拟得到了周期和长半轴的关系，并与实验拟合。

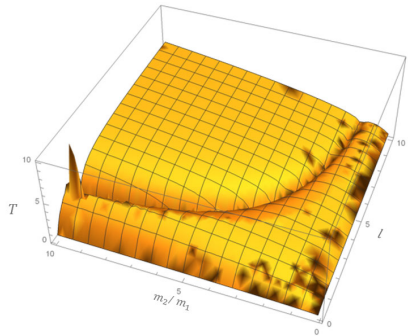

图 16 周期与绳子和质量比例的关系

如图 16 所示,可以看出周期 T 与绳长 a 和质量比例 $\frac{m_2}{m_1}$ 呈现二元函数关系。

当绳长为 0.4 时 周期与绳长的曲线如图 17 所示。当质量比例为 1 时 周期与质量比例的曲线如图 18 所示。

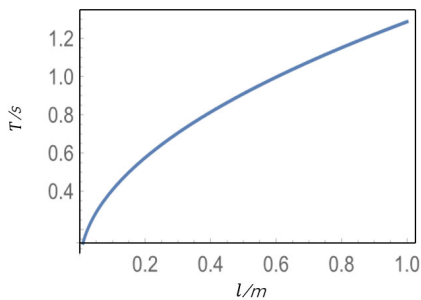 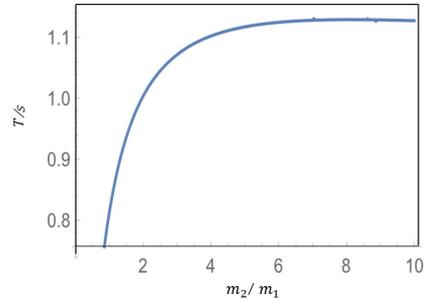

图 17 周期与绳子长的关系　　　　　图 18 周期与质量关系的关系

结论 4:周期 T 与绳长 a 和质量比例 $\frac{m_2}{m_1}$ 为二元函数关系。

## 4　结论

综上所述,本文建立了两种小球轨迹模型,给出滞后摆的轨迹经验公式、天体摆质量比、振幅、纵横比和周期等参数之间的关系。利用双目测定三维捕捉装置对绳子的自由端和两小球轨迹进行捕捉,测得小球实验运动轨迹。通过对实验轨迹与理论计算轨迹的拟合,验证了本文理论的正确性。依据此理论模型,可推测小球运动轨迹以及改变小球运动轨迹的参数。

本实验为 2020 年中国大学生物理学术竞赛题目,通过实验理论的研究机数据的拟合,获得全国二等奖成绩。并且本实验的实验现象明显,测量方法简便,可以作为大学物理设计性实验在实验室中开放,它不仅可以激发学生对物理学习的热情,也可以培养学生理论与实践相结合的综合科研能力。

## 参考文献(References)